\documentclass[11pt]{article}
\usepackage{moriond,epsfig}
\usepackage{url}

\def\be{\begin{equation}}
\def\ee{\end{equation}}
\def\bea{\begin{eqnarray}}
\def\eea{\end{eqnarray}}

\newcommand{\C}{\mathcal{C}}

\begin{document}
\vspace*{4cm}
\title{A RELATIVISTIC AND AUTONOMOUS NAVIGATION SATELLITE SYSTEM}

\author{ P. DELVA }

\address{SYRTE, Observatoire de Paris, CNRS and UPMC,\\
61 avenue de l'Observatoire, F-75014 Paris, France
}

\author{ A. \v{C}ADE\v{Z}, U. KOSTI\'{C} }

\address{Department of Physics, University of Ljubljana\\
Jadranska 19, 1000 Ljubljana, Slovenia
}

\author{ S. CARLONI }

\address{ESA Advanced Concepts Team, ESTEC, DG-PI\\
Keplerlaan 1, 2201 AZ Noordwijk, The Netherlands
}

\maketitle\abstracts{
A relativistic positioning system has been proposed by Bartolom{\'e} Coll in 2002.
Since then, several group developed this topic with different approaches. I will
present a work done in collaboration with Ljubljana University and the ESA
Advanced Concepts Team. We developed a concept, Autonomous Basis of Coordinates,
in order to take advantage of the full autonomy of a satellite constellation for
navigation and positioning, by means of satellite inter-links. I will present
the advantages of this new paradigm and a number of potential application for
reference systems, geophysics and relativistic gravitation.
}

\section{Relativistic Positioning Systems (RPS)} 

The first proposal for a relativistic positioning system is SYPOR (``Syst{\`e}me
de Positionnement Relativiste''), proposed by Bartolom{\'e} Coll in
2002~\cite{coll03a}. It is an alternative to the scheme of usual positioning
systems.  The idea is to give the constellation of satellites the possibility to
constitute by itself a \emph{primary} and \emph{autonomous} positioning system,
without any \emph{a priori} realization of a terrestrial reference frame.

The relativistic positioning system is defined with the introduction of
\emph{emission coordinates}, which contain dynamical information of the
satellite constellation. They have been reintroduced recently by several
articles~\cite{coll91,rovelli02,blagojevic02,Lachieze-Rey2006}. The definition
of these coordinates is rather simple, but they are a very powerful tool in
general relativity. Let us define four particles $a=1,2,3,4$ coupled to general
relativity. Along their worldlines $C_a$, one defines four one-parameter
families of future null cones $N_a(\tau^a)$ which are parametrized by proper time.
The intersection of four future null cones $N_a(\tau^a)$ from four worldlines $\C_a$
defines an event with emission coordinates $(\tau^1,\tau^2,\tau^3,\tau^4)$.
Then, a user receiving four electromagnetic signals broadcasting the proper time
of four satellites knows its position in this particular coordinate system.

RPS have been studied with different approaches these last
years~\cite{bahder01,coll03b,coll06b,coll06c,pascualsanchez07a,ruggiero08,bini08,tarantola09,coll2009,tartaglia10a,cadez10,coll2010,coll2010a,tartaglia11,cadez11,delva11}.
The purpose of this proceeding is to present an approach developed by a
collaboration between the Advanced Concepts Team of the European Space Agency,
the department of physics of Ljubljana University and SYRTE/Paris
Observatory/UPMC.

\section{From emission to global coordinates}

A GNSS is a system of satellites emitting precise timing signals for the purpose
of providing a local coordinate basis in space-time. In order to
determine his space-time position with respect to this basis, an observer must
receive four proper times emitted by four different satellites, and be able to
calculate the local coordinates of the four satellites as a function of
their emission coordinates. 

We studied the use of emission coordinates theoretically and in practical
scenarii by numerical simulations~\cite{cadez10,cadez11,delva11}.  We provided
mathematical tools to translate emission coordinates into space-time coordinates
of the observer. A local Schwarzschild coordinate system was introduced as an
idealized prototype of space-time in the vicinity of the Earth. The problem of
connecting the local Schwarzschild frame to the global inertial frame is well
understood in the framework of classical non-relativistic gravitational
perturbation theory, but remains to be done in a general relativistic framework.

Analytic solutions for light-like and time-like geodesics were obtained in order
to implement two algorithms: (i) an algorithm that calculates emission
coordinates corresponding to the local Schwarzschild coordinates of a user, and
(ii) the ``reverse'' algorithm that calculates space-time coordinates of a user
from its emission coordinates. In a first approach, we assumed that orbital
parameters of satellites are known. We have shown that the use of a fully
relativistic code in GNSS offers a very promising alternative to the use of
post-Newtonian approximations, and presents no technical obstacle.

The effects of non-gravitational perturbations have been studied. We have shown
that the only yardstick of a GNSS is the clock, which provides absolute position
both in space and time to any accuracy and stability allowed by noise and clock
drifts. Clock drifts, adding up after some time, would result in considerable
error in absolute position in space if the the clocks were not controlled. In
current positioning scheme, correcting the clocks needs a constant monitoring of
the satellites via Earth telemetry and the precise realization of a terrestrial
reference frame. However, we realized that a GNSS constellation is also a very
precise clock of its own, since orbital periods of its satellites are accurate
constants of motion. Therefore, we proposed to use the dynamical information
given by mutual timing between satellites to improve the long term phase
stability of onboard clocks, as well as to improve the precision of constants of
motion of the constellation. This proposal led us to define the concept of
Autonomous Basis of Coordinates~(ABC). Within such a scheme, we have shown that
it is possible to correct the clocks to a level considerably surpassing the
classical scheme, which is limited in accuracy by stochastic components of Earth
dynamics. 

\section{Autonomous Basis of Coordinates (ABC)}

In a GNSS constellation with more than four satellites, more than four emission
coordinates are received by an observer: the positioning problem is
over-determined. In order for the local basis to be self consistent, all
combinations of emission coordinates, received at any event in space-time, must
give the same four local coordinates for this event. The main constraint on
self-consistency of a GNSS system comes from the precision of constants of
motion. In order to adress this problem, the concept of Autonomous
Basis of Coordinates~(ABC) is introduced in \v{C}ade\v{z} \emph{et
al.}~\cite{cadez11}. We propose that the constants of motion be determined and
checked internally by the GNSS system in such a way that each satellite checks
its own position as any other observer with respect to all the other satellites:
in addition to emitting its proper time, each satellite also receives other
satellite's emission coordinates and makes its information available to the
central GNSS control~\cite{amarillo,gnssplus}. 

The ABC concept aims to describe in a coherent frame both the dynamics of
non-interacting test bodies transmitting emission coordinates and the
propagation of electromagnetic waves providing those coordinates. It uses the
fact that both light and test bodies trajectories are geodesics that can be
derived from the same Hamiltonian. It provides a means to translate dynamical
information into the conventional representation based on local frames.
Dynamical information, expressed in terms of emission coordinates, gives direct
information about the Riemannian structure of space-time, and thus allows the
construction of a local frame with coordinates and metric that provides a
precise definition of equations of motion. We call the reference system and
coordinates built via the ABC concept the \emph{ABC reference system} and the
\emph{ABC coordinates}.

Let us use the nomenclature introduced by J.~Kovalevsky and
I.~Mueller~\cite{kovalevsky89} to describe the ABC reference system:

\noindent {\bf Concept:} 
the ABC coordinate system is built such that dynamics is consistent; dynamics
is given by a Hamiltonian, that both describe space-time geometry
and non-gravitational forces.

\noindent {\bf Physical structure:} the reference system is physically materialized
by a constellation of satellites in Earth orbit and inter-satellite links.
Light and satellite geodesics create a physical space-time web that probe the
space-time geometry.

\noindent {\bf Modelling:} the model characterizes a particular choice of the
Hamiltonian. We have studied three particular Hamiltonians~\cite{cadez11}:
Minkowski, Kepler and Schwarzschild. The ultimate goal is to obtain a
Hamiltionian containing a complete description of all known gravitational
and non-gravitational perturbations. This is the purpose of the
Slovenian PECS/ESA project: ``Relativistic global navigation
system''~\footnote{\url{http://www.esa.int/SPECIALS/PECS/index.html}}.

\noindent {\bf Realization:} A realization of the reference system needs the
implementation on future GNSS constellation of inter-satellite links, which is
now under study~\cite{amarillo,gnssplus}. We have done a
simulation~\cite{cadez10,cadez11,delva11} for some specific idealized space-time
geometries and have discovered some generic properties of ABC systems, as
robustness of recovering constants of motion with respect to noise in the data,
consistency of description with redundant number of satellites, the possibility
to use the constellation as a clock with long term stability and the possibility
to use perturbation theory to refine the Hamiltonian toward a better long term
dynamical prediction. For example, we have shown an internal consistency of
Galileo satellites positions at the millimetre level after only four orbits
($\sim 36$ hours) with 200 data points (one point every 10~mn). The accuracy of
constants of motion is expected to increase with time, when more data will
become available to evaluate smaller and smaller discrepancies between dynamic
prediction and dynamic observations provided by exchange of emission coordinates
between satellites.


\section{Applications in geophysics and relativistic gravimetry}

Dynamics of bodies and light in a given space-time is unique to the geometry of
this space-time. Therefore, geometry can in principle be determined on the basis
of dynamical information and vice versa, dynamics can be predicted with an
accuracy limited, in principle, only by the accuracy of geometric information.
Thus the GNSS with inter-satellite links is a new type of gravimeter, we call it
\emph{Riemannian gravimeter}, that creates a space-time web with light and
satellite geodesics that ``scan'' the space-time geometry around Earth.

The accuracy of an ABC reference system, realized with Galileo satellites, would
increase with the accuracy of geometric information derived from dynamics.
However, the relation of such an ABC reference system to a celestial reference
system is not trivial, since the ABC reference system is gauged with the local
geometry of the part of space-time where satellites move, while signals from
distant quasars travel long distances accross the universe and are affected by
the intervening curvature of spacetime. Thus, the relation between the ABC
reference frame and a celestial reference frame could, in principle, reveal
important new information about the way in which the local geometry is
integrated into the global arena of space-time. A discrepancy between the two
frames could also reveal a violation of the equivalence principle, if
non-gravitational perturbations as solar pressure can be modelled or measured
accurately.


The possibility to define an extremely precise ABC reference frame is also very
interesting for geophysics. A sub-millimetre level of accuracy of satellite
positions would eventualy allow comparable position accuracy on Earth surface,
at least statistically, by properly averaging positions obtained by ground based
GNSS receivers. Below millimetre level of accuracy, the shape of Earth and
absolute positions of markers on the ground would certainly elucidate many
important phenomena about our planet Earth. For example, a much deeper
understanding of interior structure of the Earth could be reached by studying
Earth and ocean tides. Continental drift would be measured with a precision,
that could possibly be sufficient to model changing strain and stress in the
Earth crust and eventually lead to earthquake prediction. Gravitational
potential differences and driving ocean currents could also be detected,
allowing us to study ocean dynamics at the same level of precision as todays
meteorology understands dynamics of atmosphere.




\section*{References}

\begin{thebibliography}{10}
\providecommand{\url}[1]{\texttt{#1}}
\providecommand{\urlprefix}{URL }

\bibitem{coll03a}
B.~Coll, in N.~Capitaine and M.~Stavinschi, eds., \emph{Journ{\'{e}}es 2002 -
  syst{\`{e}}mes de r{\'{e}}f{\'{e}}rence spatio-temporels. Astrometry from
  ground and from space, Bucharest, 25 - 28 September 2002} \textbf{14}, 34 (2003).

\bibitem{coll91}
B.~Coll and J.~A. Morales, \emph{Journal of Mathematical Physics} \textbf{32},
  9 2450 (1991).

\bibitem{rovelli02}
C.~{Rovelli}, \emph{Phys. Rev. D} \textbf{65}, 4 044017 (2002).

\bibitem{blagojevic02}
M.~{Blagojevi{\'c}} \emph{et~al.}, \emph{Phys. Rev. D} \textbf{65}, 4 044018
  (2002).

\bibitem{Lachieze-Rey2006}
M.~Lachi\`{e}ze-Rey, \emph{Class. Quantum Grav.} \textbf{23}, 10 3531 (2006).

\bibitem{bahder01}
T.~B. {Bahder}, \emph{Am. J. Phys.} \textbf{69} 315 (2001).

\bibitem{coll03b}
B.~Coll and A.~Tarantola, poster at JSR 2003,
  ``Astrometry, Geodynamics and Solar System Dynamics: from milliarcseconds to
  microarcseconds'', St. Petersburg, September 22 - 25, 2003 (2003),
  \urlprefix\url{http://syrte.obspm.fr/~coll/}.

\bibitem{coll06b}
B.~{Coll} \emph{et~al.}, \emph{Phys. Rev. D} \textbf{73}, 8 084017 (2006).

\bibitem{coll06c}
B.~{Coll} \emph{et~al.}, \emph{Phys. Rev. D} \textbf{74}, 10 104003 (2006).

\bibitem{pascualsanchez07a}
J.-F. Pascual-S\'{a}nchez, \emph{Ann. Phys. (Leipzig)} \textbf{16} 258 (2007).

\bibitem{ruggiero08}
M.~L. {Ruggiero} and A.~{Tartaglia}, \emph{Int. J. Mod. Phys. D} \textbf{17}
  311 (2008).

\bibitem{bini08}
D.~Bini \emph{et~al.}, \emph{Class. Quantum Grav.} \textbf{25}, 20 205011
  (2008).

\bibitem{tarantola09}
A.~{Tarantola} \emph{et~al.}, \emph{ArXiv e-prints}  (2009).

\bibitem{coll2009}
B.~Coll \emph{et~al.}, \emph{Phys. Rev. D} \textbf{80}, 6 064038 (2009).

\bibitem{tartaglia10a}
A.~Tartaglia, \emph{Acta Astronautica} \textbf{67} 539 (2010).

\bibitem{cadez10}
A.~\v{C}ade\v{z} \emph{et~al.}, {Mapping the Spacetime Metric with a Global
  Navigation Satellite System - Final Report}, Tech. rep., Advanced Concepts
  Team, European Space Agency (2010).

\bibitem{coll2010}
B.~Coll \emph{et~al.}, \emph{Classical and Quantum Gravity} \textbf{27}, 6
  065013 (2010).

\bibitem{coll2010a}
B.~Coll \emph{et~al.}, \emph{Phys. Rev. D} \textbf{82}, 8 084038 (2010).

\bibitem{tartaglia11}
A.~Tartaglia \emph{et~al.}, \emph{Advances in Space Research} \textbf{47} 645
  (2011).

\bibitem{cadez11}
A.~\v{C}ade\v{z} \emph{et~al.}, {Mapping the Spacetime Metric with a Global
  Navigation Satellite System - Extension of study: Recovering of orbital
  constants using inter-satellites links}, Tech. rep., Advanced Concepts Team,
  European Space Agency (2011).

\bibitem{delva11}
P.~Delva \emph{et~al.}, \emph{Advances in Space Research} \textbf{47} 370
  (2011).

\bibitem{amarillo}
I.~{Rodr{\'{\i}}guez-P{\'e}rez} \emph{et~al.}, \emph{Advances in Space
  Research} \textbf{47} 197 (2011).

\bibitem{gnssplus}
L.~{Arona} \emph{et~al.}, {GNSSPLUS, Final Report.
  GNSSPLUS-DMS-TEC-FIR01-11-E-R} (2006).

\bibitem{kovalevsky89}
J.~{Kovalevsky} \emph{et~al.}, eds., \emph{{Reference frames in astronomy and
  geophysics}}, \textbf{154} (1989).

\end{thebibliography}


\end{document}